\documentclass[conference]{IEEEtran}
\IEEEoverridecommandlockouts

\usepackage{epsfig} 
\usepackage{gensymb}
\usepackage{graphicx, color, soul}
\usepackage{amsmath, amsfonts, amssymb}
\usepackage[in]{fullpage}
\usepackage[small, bf]{caption}
\usepackage[normalem]{ulem}
\usepackage[margin=1in]{geometry}
\usepackage{floatrow}
\usepackage{multirow}
\newfloatcommand{capbtabbox}{table}[][\FBwidth]
\usepackage{biblatex}
\usepackage{booktabs}

\newcommand{\fig}[1]{Fig.~\ref{figure:#1}}

\newcommand{\figur}[1]{Figure~\ref{figure:#1}}

\newcommand{\eq}[1]{Eqn.~(\ref{equation:#1})}

\newcommand{\be}{\begin{equation}}
\newcommand{\ee}{\end{equation}}
\newcommand{\bea}{\begin{eqnarray}}
\newcommand{\eea}{\end{eqnarray}}
\newcommand{\beau}{\begin{eqnarray*}}
\newcommand{\eeau}{\end{eqnarray*}}
\newcommand{\bed}{\begin{displaymath}}
\newcommand{\eed}{\end{displaymath}}

\begin{document}


\title{A Lightweight Algorithm for Classifying Ex Vivo Tissues Samples\\

\thanks{This work was supported in part by the U.S. National
Science Foundation (grant No. 2221174) and the U.S. National Institutes of Health (grant Nos. 1R01CA237654-01A1 and 1R21DE031095-01).}
}


\author{
    \IEEEauthorblockN{ Tzu-Hao Li, Ethan Murphy, Allaire Doussan, Ryan Halter, Kofi Odame}
    \IEEEauthorblockA{Thayer School of Engineering at Dartmouth College, Hanover NH, USA}
}

\maketitle

\begin{abstract}
In this paper, we present a novel algorithm for classifying ex vivo tissue that comprises multi-channel bioimpedance analysis and a hardware neural network. When implemented in a mixed-signal 180 nm CMOS process, the classifier has an estimated power budget of 39 mW and an area of 30 mm$^2$. This means that the classifier can be integrated into the tip of a surgical margin assessment probe, for in vivo use during radical prostatectomy. We tested our classifier on digital phantoms of prostate tissue and also on an animal model of ex vivo bovine tissue. The classifier achieved  an accuracy of 90\% on the prostate tissue phantoms, and an accuracy of 84\% on the animal model.
\end{abstract}

\section{Introduction}

Localized prostate cancer is typically treatable with radical prostatectomy, which presents a  tradeoff between conserving peri-prostatic tissue (to avoid urinary and erectile dysfunction) and leaving cancerous tissue behind (i.e. a positive surgical margin). When there is a positive surgical margin, there is a significant risk of disease recurrence \cite{preisser2019rates}, which would necessitate radiotherapy or hormonal treatments with adverse side-effects like cystitis, rectal bleeding and cognitive impairment. 

To avoid the negative outcomes associated with a positive surgical margin (PSM), we have been developing bioimpedance (`bioZ’) analysis for in vivo  surgical margin assessment during radical prostatectomy \cite{murphy2017comparative}. The challenge of a bioZ system is that, to minimize signal loss and cross-talk, the analog front end (AFE) must be physically close to the electrode array \cite{kim20171}. Achieving such physical proximity in an in vivo application would require the AFE and electrode array to fit inside a millimeter-sized endoscope-type surgical margin assessment probe \cite{doussan2023towards}.

\begin{figure}
\begin{center}
\includegraphics[scale=0.63]{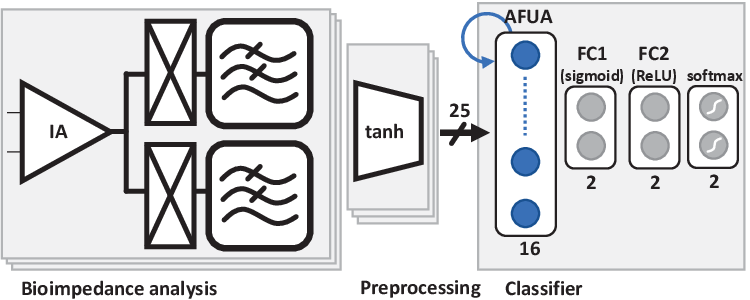}
\end{center}
\caption{Proposed approach for prostate cancer detection involves 25-channel
bioZ analysis, followed by a preprocessing stage and then a neural network. The neural network has the following architecture: Layer 1: 25-unit input stage; Layer 2: 16 units of a type of long short-term memory (AFUA); Layer 3: 2-unit fully-connected layer with sigmoid activation (FC1); Layer 4: 2-unit fully-connected layer with ReLU activation (FC2); Layer 5: 2-unit softmax output layer.}
\label{figure:mainArgument}
\end{figure}

Unfortunately, such a small probe cannot accommodate conventional bioZ analog front ends (AFEs), which contain several large analog-to-digital converters (ADCs). To design a suitably small AFE, we can omit the large ADCs and instead directly analyze the bioZ signals in the analog domain. In this paper, we introduce a lightweight classifier for prostate cancer detection that can be integrated into a bioZ analog front end.

\begin{figure*}[ht]
\begin{center}
\includegraphics[scale=0.86]{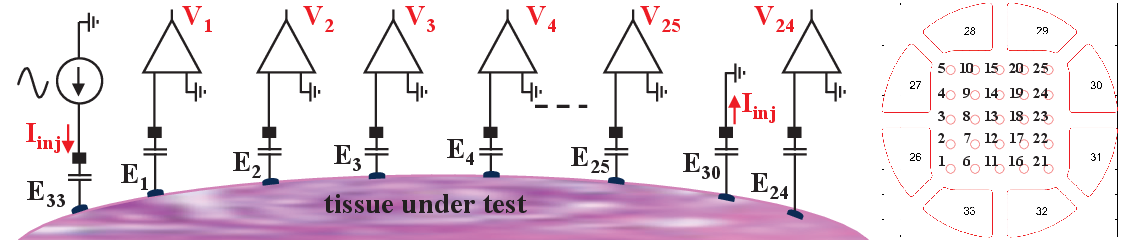}
\end{center}
\caption{Left panel: Example bioZ measurement. One set of bioZ data is collected by injecting a 1 mA$_{\rm pp}$ sinusoidal current ($I_{\rm inj}$) via a pair of electrodes (E33, E30) and measuring the induced voltages (V1, V2, etc.) with instrumentation amplifiers. These voltages are then processed to extract their amplitude and phase components. Voltage measurements are repeated for all possible current sink/source electrodes pair combinations (or `current patterns') to form a `frame' of bioZ data. Right panel: Electrode array configuration for our proposed surgical margin assessment probe. The voltage measurement electrodes are the small inner electrodes (i.e. electrodes 1 to 25). The current injection/sink electrodes are the larger electrodes on the probe's outer ring (i.e. electrodes 26 to 33). There are 8-choose-2=28 possible current patterns.}
\label{figure:bioZ}
\end{figure*}

\section{Algorithm Overview}
Our approach for prostate cancer detection involves  bioZ analysis, followed by a preprocessing stage and then a neural network (see \fig{mainArgument}). The preprocessing stage normalizes the raw single-ended voltages produced by bioZ analysis of the tissue. After normalization, the single-ended voltages are processed by a long short-term memory (LSTM) neural network that labels the tissue as either benign or as containing a cancerous lesion.

\subsection{Bioimpedance Analysis}
To perform bioZ analysis, we pass an imperceptibly small sinusoidal current through a prostate tissue sample via two current electrodes (known as a `current pattern’). As the current passes through the tissue, it produces an electric potential on the tissue surface, which we  measure with a set of voltage electrodes,  \fig{bioZ}. We repeat this measurement for all possible current patterns (i.e. all possible pairs of current sink/source electrodes) to create a sequence or `frame' of bioZ data. The frame is processed to determine if the tissue under test contains a cancerous lesion.

\subsection{Preprocessing}
The preprocessing stage normalizes each frame by first taking the difference between each single-ended voltage (measured from the tissue under test) and a corresponding reference voltage (measured from a saline solution). Then, the differenced voltage is input to a $\tanh$ function to constrain it to values between $\pm$1. The preprocessing stage can be described with the function
\begin{equation}
\mathbf{x} = \tanh\left( |V_{\rm meas}| - |V_{\rm ref}| \right),
\label{equation:preproc}
\end{equation}
where $V_{\rm meas}$ is the frame of measured single-ended voltages and $V_{\rm ref}$ is the reference frame.

\subsection{LSTM Neural Network Architecture}
 Compared to alternatives like convolutional neural networks and transformers, LSTMs do not need internal data buffers, and so can be implemented in a small area. In this paper, we use a modified version of our recently introduced \emph{adaptive filter unit for analog LSTM} (AFUA) \cite{odameAFUAEat, odameAFUAKWS}, that can be efficiently implemented as an analog circuit:
\begin{eqnarray}
z_j(t) & = & f([\mathbf{W}_z \mathbf{x}]_j + [\mathbf{U}_z (\mathbf{h}(t))]_j) \nonumber \\ 
\tilde{h}_j(t) & = & f([\mathbf{W} \mathbf{x}]_j + [\mathbf{U}(\mathbf{h}(t))]_j) \nonumber \\ 
\frac{\tau_h}{z_j(t)}\frac{dh_j}{dt} & = & \left(1 - \frac{h_j(t)}{\tilde{h}_j(t)}\right),
\label{equation:AFUA}
\end{eqnarray}
where $x$ is the input, $h_j$ is the hidden state, $\tilde{h}_j$ is the candidate state, $z_j$ is the update gate and $W_{*}$ and $U_{*}$ are learnable weight matrices. Also, $f(\cdot)$ is a sigmoid-type activation function with a range of $(0, 1)$. 

Like a conventional LSTM, the AFUA is able to selectively retain or update its memory of input data. But while the LSTM contains 3 Hadamard multiplications, the AFUA contains none. Also, since \eq{AFUA} can be implemented as a continuous-time current-mode circuit, the AFUA avoids the overhead of operational amplifiers, current/voltage converters and internal digital/analog converters found in other analog LSTM implementations \cite{li2021ns}. Fewer operations and smaller overhead make the AFUA extremely hardware efficient and suitable for integration into the constrained space of a surgical margin assessment probe.

Our AFUA neural network comprises a 25-unit input stage and a 16-unit AFUA layer. Following the AFUA layer is a 2-unit fully connected layer with sigmoidal activation function, a second 2-unit layer with fully-connected layer with ReLU activation function, and finally a softmax output layer (see \fig{mainArgument}).

\section{Experimental Approach}

\subsection{Digital Phantom}
\label{section:digital}
To validate our prostate cancer detection approach, we constructed digital phantoms of ex vivo prostate tissue from a 5 mm thick 3D finite element method (FEM) mesh with the probe's electrode array configuration (see \fig{bioZ}) encoded on the top surface. Based on ex vivo values at 10 kHz, we modeled normal background tissue with a conductivity of $126+12.76j$ mS/m, while cancerous regions had a conductivity of $106+14.9j$ mS/m. Noise was added to the background via a radial-basis-function neural network, which was tuned to produce an overall standard deviation of background noise to be 10\%---similar to what we might expect from real tissue. 

We created 4265 digital phantoms of prostatic tissue slices that each contained an inclusion of cancerous region, ranging in diameter from 0 mm to 3 mm, and located in a random part of the phantom. If a phantom contained an inclusion of area 8\% or more of the probe, then we considered this phantom to be positive for cancer. Otherwise, it was labeled negative. 
We simulated bioZ analysis at the center of each phantom to produce a synthetic frame of $28\times25$ single-ended voltages (that is, 25 voltage electrode measurements for each of the 8-choose-2 = 28 current injection patterns).

\subsection{Animal Model}
\begin{figure}[t]
\includegraphics[scale=0.24]{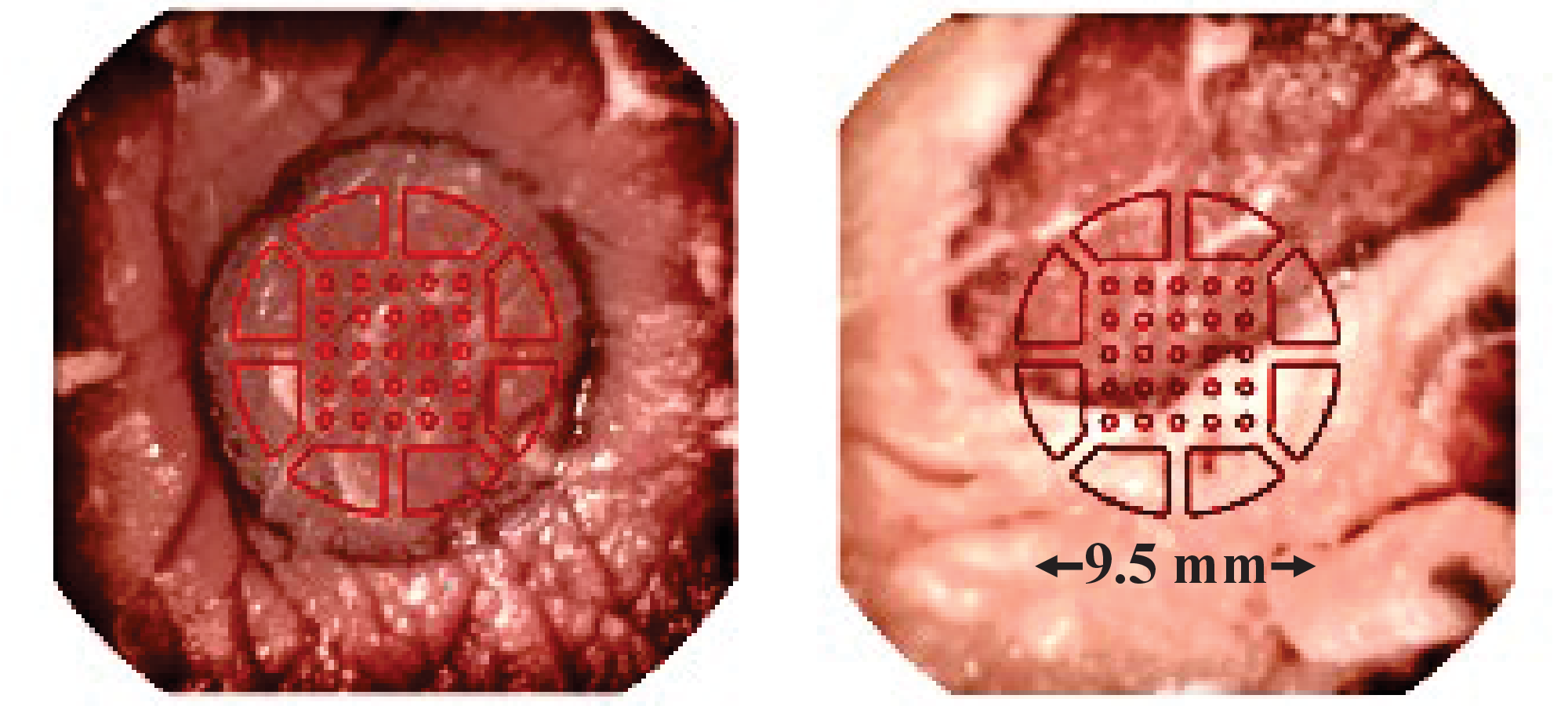}
\caption{Images of ex vivo bovine tissue samples with (left panel) mostly muscle tissue and (right panel) mixture of muscle and adipose tissue. The electrode array overlay (shown in red) indicates where a bioimpedance measurement was taken.
}
\label{figure:exvivo}
\end{figure}

We recorded bioZ measurements from an animal model that comprised ex vivo bovine tissue containing a mixture of muscle and adipose (see \fig{exvivo}), using the data collection protocol described in a previous study \cite{doussan2023towards}. In this ex vivo bovine model, the muscle tissue represents the background (analogous to normal prostate tissue), while the adipose tissue represents the feature of interest (analogous to cancerous lesion in prostate tissue). Although the adipose-muscle contrast is higher than that of prostatic cancerous-normal tissue, this animal model allowed us to do some early tests of our classifier on real-world data.

Taking bioZ measurements from tissue slices is a lengthy process, and only a small amount of physical data could be collected in a practical amount of time. To have enough data for neural network training, we generated synthetic bioZ measurements from digital phantoms of ex vivo bovine tissue (see Sect. \ref{section:digital}), modeling background muscle tissue with a conductivity of $341 + 14.4j$ mS/m, and adipose tissue with a conductivity of $23.8 + 0.604j$ mS/m \cite{Andreuccetti}. We simulated bioZ measurements at the center of each of 2988 digital phantoms, to generate 2988 frames of synthetic 28$\times$25 single-ended voltages for training the ex vivo bovine classifier.

\subsection{Experiments}

\subsubsection{Prostate Digital Phantom}
We tested our classifier on the prostate digital phantom data, using 2398 samples for training, 800 for validation and 1067 held out for testing. We trained the neural network in Python with a batch size of 100 for 500 epochs. The resulting learned weights were  converted from floating-point representation to signed fixed point representation to study the effect of quantization; weight quantization is critical for efficient hardware implementation, but it typically causes a drop in classification accuracy. The quantized models were tested on the held out set.

\subsubsection{Ex Vivo Bovine Model}
In order to test our classifier on the animal model, we first trained and validated it on the synthetic data that was generated from the 2988 digital phantoms of ex vivo bovine tissue. We separated the synthetic data into two sets: 2241 for training and 747 for validation. We performed weight quantization (conversion to 4-bit signed fixed point representation) and tested the final, quantized model on the bioZ data that was measured from real ex vivo bovine tissue.

\subsubsection{Hardware Analysis}
\begin{figure}
\begin{center}
\includegraphics[scale=0.3]{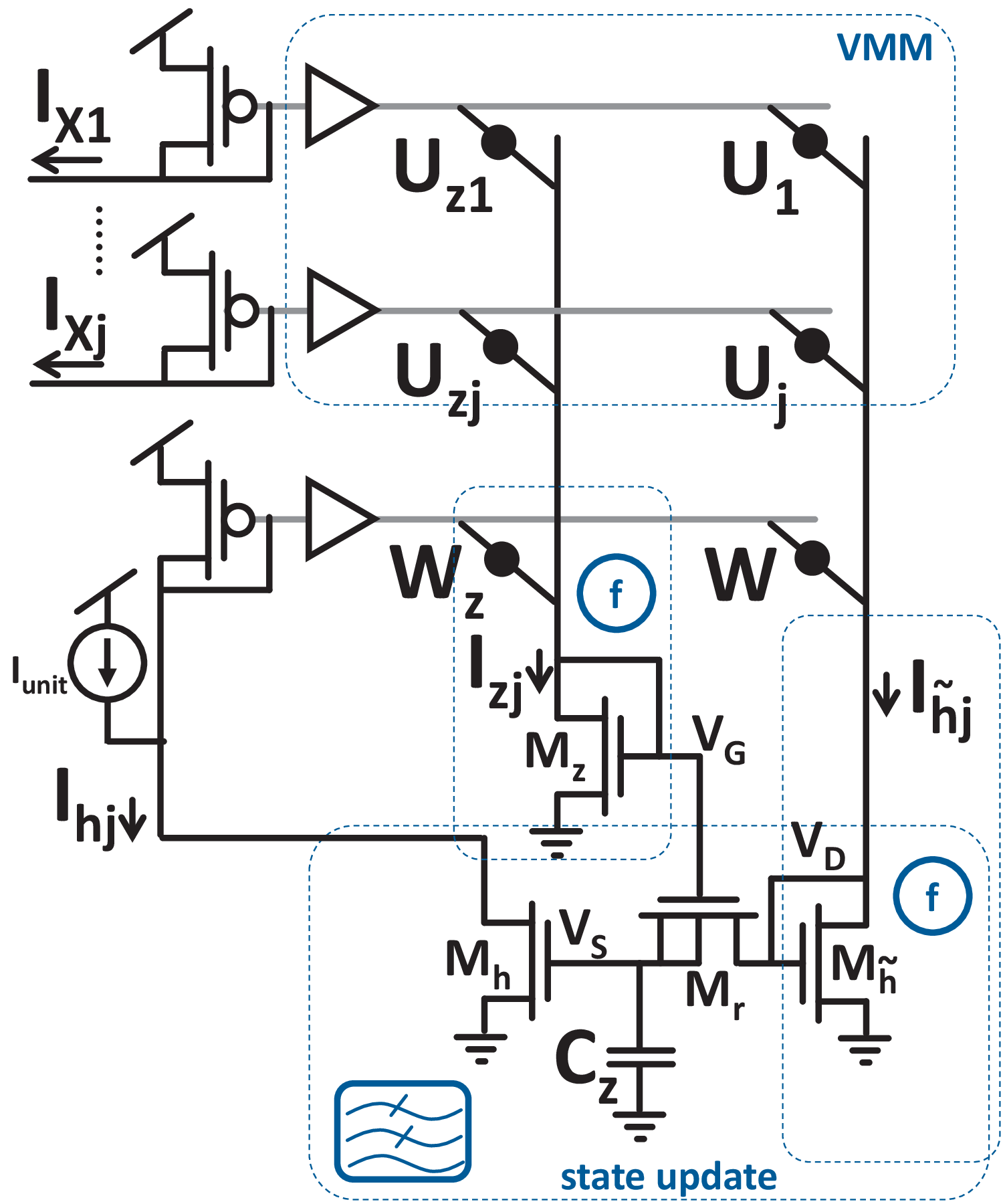}
\end{center}
\caption{AFUA schematic, comprising vector matrix multiplier (`VMM') and state update implemented as a gated current mirror. The  current mirror incorporates  activation functions that consume no extra power. The VMM is implemented using standard current mirror-based circuitry \cite{binas2016precise}.}
\label{figure:AFUAckt}
\end{figure}

To assess the hardware requirements of our classifier, we designed the AFUA neural network as an application specific integrated circuit (ASIC) in a 180 nm CMOS process (\fig{AFUAckt}). From the schematic, we can apply Kirchhoff's current law to the gate of $M_h$ to find the drain current of $M_h$
(assuming subthreshold operation) to be
\begin{equation}
\tau_h \frac{dI_{hj}}{dt} = I_{zj}\left(1 - \frac{I_{hj}}{I_{\tilde{h}j}}\right),
\label{equation:gruprUpdate}
\end{equation}
where $\tau_h$ is a time constant that depends on the size of capacitor $C_z$. \eq{gruprUpdate} is the current-mode representation of the AFUA state update function in \eq{AFUA}. 

We analyzed the neural network ASIC in the Virtuoso ADE Suite (Cadence Design Systems Inc., San Jose, CA) circuit simulator to determine the power and area requirements of our proposed approach.

\section{Results and Discussion}

\subsubsection{Prostate Digital Phantom}
\begin{figure}
\includegraphics[scale=0.54]{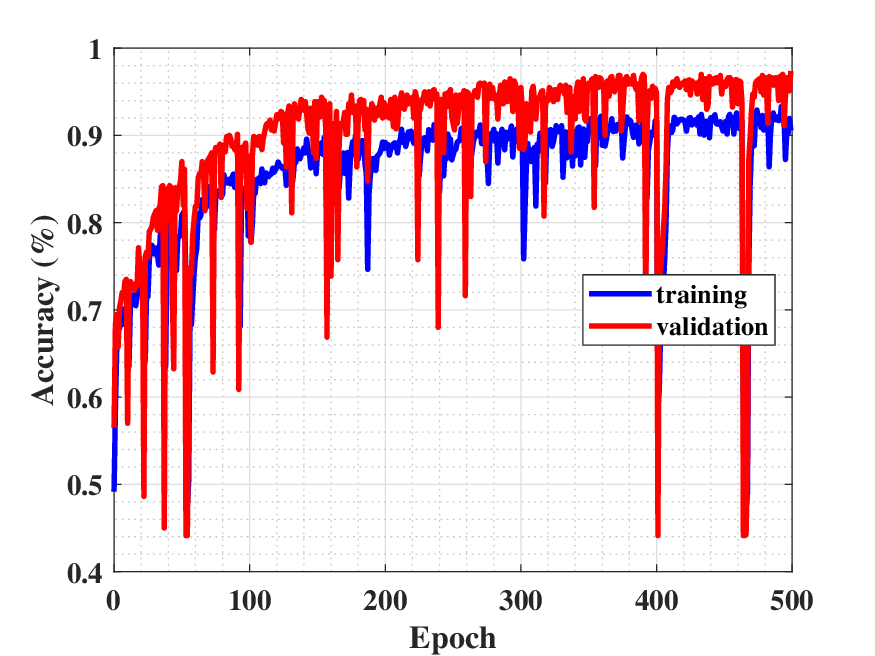}
\caption{Prostate digital phantom task. Accuracy curves for the full-precision (64-bit floating point) neural network model when evaluated on the training and validation sets of the prostate digital phantom data.}
\label{figure:training}
\end{figure}

\begin{figure}
\includegraphics[scale=0.54]{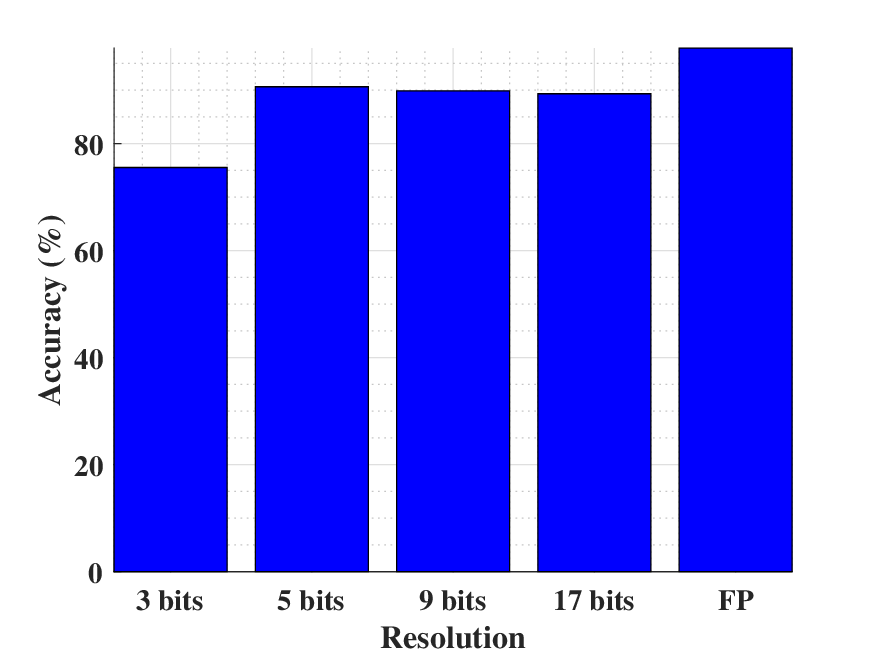}
\caption{Effect of quantization on the prostate digital phantom task. Classification accuracy is at least 90\% if the model is quantized at 5-bit resolution or higher. \emph{N bits} denotes signed (N-1)-bit  fixed point representation of the neural network weights. \emph{FP} denotes full precision, 64-bit floating point representation.}
\label{figure:bits}
\end{figure}

\figur{training} shows that the classifier (with full-precision floating point weights) is able to train effectively on the prostate digital phantom data, with no indication of overfitting. When evaluated on the held-out test set of prostate digital phantom data, the full-precision classifier performed with an accuracy of 97.8\%. From \fig{bits}, we see that quantization causes classification accuracy to drop to 90\% for resolutions of 5 bits (i.e. signed 4-bit  fixed point representation) or higher. But accuracy drops to 75\% when the neural network weights are quantized with 3-bit resolution.

\subsubsection{Ex Vivo Bovine Model}
\begin{figure}
\begin{center}
\includegraphics[scale=0.5]{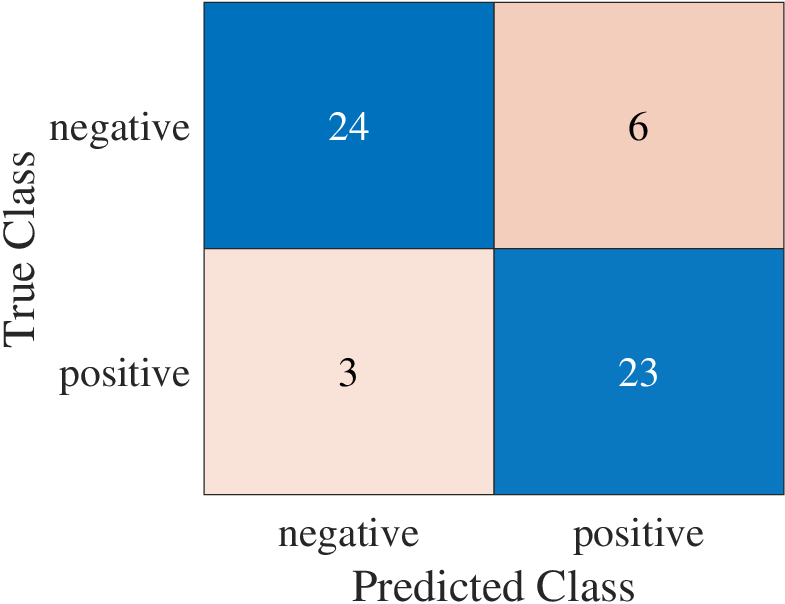}
\end{center}
\caption{Confusion matrix of 5-bit quantized classifier, tested on the dataset measured from the animal model of ex vivo bovine tissue. Ground truth (`True Class') is labeled positive for fat if the probed location contains 8\% or more of adipose tissue. Otherwise, it is labeled negative.}
\label{figure:confusion}
\end{figure}

The classifier was similarly able to train on the ex vivo bovine digital phantom data, and the 5-bit quantized version of the classifier yielded a validation set accuracy of 90\%. As the confusion matrix of \fig{confusion} shows, the accuracy dropped to 84\% when the classifier was evaluated on bioZ measurements of real ex vivo bovine tissue. The drop in accuracy is due to non-idealities in the bioZ measurement, including variations in contact pressure and poor electrode contact.

\subsubsection{Hardware Analysis}
The majority of the chip area is taken by the vector matrix multipliers (VMMs) in the AFUA and fully-connected layers (\fig{AFUAckt}). A VMM comprises an array of programmable current sources \cite{odameAFUAEat}, each of which has an area of 220 $\mu$m$\times$40 $\mu$m to meet analog matching requirements. Between the AFUA layer VMM and those of the fully-connected layers, there are 2592 programmable current sources. So, including an allowance for routing and other circuit components, we estimate a total chip area of 30 mm$^2$.

Although the VMMs are physically large, they consume less than 100 $\mu$A, because their processing speed is only required to meet the bioZ measurement frame rate (typically 20 FPS). In contrast, each of the instrumentation amplifiers that processes the 25 voltage electrode outputs must operate at  the interrogation current frequency (10 kHz in our application), in addition to providing large differential gain, high common mode rejection ratio and low noise. As such, the instrumentation amplifiers consume the vast majority of the ASIC's supply current. Based on our previous work \cite{takhti2019power}, we estimate that the 25 instrumentation amplifiers consume a total of 11.75 mA (or 39 mW on a 3.3 V power supply).

\section{Conclusion}
In this paper, we proposed an approach to prostate cancer detection that is based on neural network processing of bioimpedance measurements. The approach yielded an accuracy of 90\% when tested on digital phantoms of ex vivo prostate tissue, and an accuracy of 84\% when tested on an animal model of ex vivo bovine tissue. We estimate that our approach can be implemented as an ASIC with an area of 30 mm$^2$ and a power consumption of 39 mW. This makes our approach a light weight solution that can be integrated into the probe tip of a surgical margin assessment tool. 

\printbibliography
\end{document}